\documentclass[
reprint,
superscriptaddress,
frontmatterverbose, 
showpacs,
preprintnumbers,
nofootinbib,
amsmath,
amssymb,
aps,
floatfix,
twocolumn
]{revtex4-2}

\usepackage{graphicx}
\usepackage{dcolumn}
\usepackage{bm}
\usepackage{hyperref}
\usepackage{xcolor}
\usepackage{subcaption}
\usepackage{booktabs,array}
\usepackage{bbm}

\newcommand{\Id}{\mathbbm{1}}
\newcommand{\Mp}{M_{\rm P}}
\newcommand{\SL}{\mathrm{SL}(2,\mathbb{Z})}

\newcommand{\gS}{\mathsf{S}}
\newcommand{\gT}{\mathsf{T}}
\newcommand{\Rc}{\mathsf{R}}
\newcommand{\Ur}{U(1)_R}

\begin{document}

\title{
Topological Inflation from a Modular Symmetry  
}

\author{Xin Wang}
 \email{xin.wang@unipd.it}
  \affiliation{Dipartimento di Fisica e Astronomia ``Galileo~Galilei'', \\
    Università degli Studi di Padova,  
    Via Francesco~Marzolo 8, 35131 Padova, Italy
 }
 \affiliation{INFN, Sezione di Padova,  
    Via Francesco~Marzolo 8, 35131 Padova, Italy
 }
\author{Tsutomu T. Yanagida}
 \affiliation{Kavli Institute for the Physics and Mathematics of the Universe (WPI), UTIAS, The University of Tokyo, Chiba 277-8583, Japan
}

\date{\today}

\begin{abstract}
We show that topological inflation arises naturally from modular symmetry in supergravity. The central element of $\mathrm{SL}(2,\mathbb{Z})$ acts as $\phi\to-\phi$ on an inflaton field $\phi$ of odd modular weight, providing the required $\mathbb{Z}_2$ symmetry for topological inflation. The fixed point $\tau=\omega\equiv e^{2\pi i/3}$ of the modulus is preferred, while $\tau = i$ fails to accommodate the minimal $\mathbb{Z}_2$ topological inflation. We numerically compute the inflationary observables and determine the parameter space consistent with the Planck and ACT results. The reheating process is also discussed, including the decay into right-handed neutrinos whose bare Majorana masses are forbidden by modular invariance.
\end{abstract}

\maketitle

\section{Introduction}
\label{sec:introduction}
The inflationary paradigm~\cite{Starobinsky:1980te,Guth:1980zm,Linde:1981mu,Albrecht:1982wi} provides a natural explanation for the horizon and flatness problems of the hot big bang cosmology. Its predictions for the primordial perturbations are supported by increasingly precise cosmological observations~\cite{Planck:2018jri,BICEP:2021xfz,AtacamaCosmologyTelescope:2025blo}. In small-field models, such as new inflation~\cite{Linde:1981mu,Albrecht:1982wi}, the inflaton starts to slow-roll near a local maximum of the potential, and eventually arrives at the minimum. However, such models usually suffer from an initial value problem, since the inflaton is required to
start extremely close to the local maximum~\cite{Izawa:1998rh,Antusch:2014qqa,Ai:2025qua}.

In this regard, topological inflation~\cite{Linde:1994hy,Vilenkin:1994pv,Linde:1994wt} is a consistent inflation model, and has been realized in supergravity in Refs.~\cite{Izawa:1998rh,Kawasaki:2000tv,Choi:2022fce,Choi:2022ssv}. In this scenario, domain walls exist before inflation, separating spatial regions in different vacua. At the core of each wall, the field is held near the local maximum by topology. If the vacuum expectation value of the field is of order the Planck scale, the wall is thicker than the Hubble radius and its core inflates~\cite{Linde:1994hy,Vilenkin:1994pv}. The domain walls thus naturally provide the initial conditions for inflation. Moreover, since these walls are pushed far beyond the present horizon, they do not pose a problem for cosmology.

Modular symmetry, which arises in string compactifications~\cite{Ferrara:1989bc,Ferrara:1989qb}, has recently been proposed to explain the flavor mixing of quarks and leptons~\cite{Feruglio:2017spp} (see Refs.~\cite{Feruglio:2019ybq,Kobayashi:2023zzc,Ding:2023htn} for reviews). Beyond flavor physics, modular symmetry may also have implications for inflation, where either the modulus field itself plays the role of the inflaton~\cite{Bento:1996xh,Kobayashi:2016mzg,Schimmrigk:2016bde,Schimmrigk:2017jwa,Lynker:2019joa,Abe:2023ylh,King:2024ssx,Ding:2024neh,Casas:2024jbw,Kallosh:2024ymt,Kallosh:2024pat,Aoki:2024ixq,Ding:2024euc,Jiang:2025qbi,Abe:2026cmt}, or modular invariance constrains the couplings of additional fields driving inflation~\cite{Gunji:2022xig,Aoki:2025wld}. Modular symmetry has also been applied to the formation of cosmological domain walls~\cite{Cvetic:1991vp,Cvetic:1992bf,Cvetic:1992cv,Cvetic:1996vr,King:2024lki}, dark matter~\cite{Nomura:2019jxj,Kobayashi:2021ajl,Baur:2024lcc} and dark energy~\cite{Anchordoqui:2025epz,Gordillo-Ruiz:2025xcg}.

In this short letter we show that the topological inflation model in Refs.~\cite{Izawa:1998rh,Kawasaki:2000tv,Choi:2022fce,Choi:2022ssv}
arises naturally from a modular symmetry. The central element $\Rc=-\Id$ of the modular group $\SL$ acts as $\phi\to-\phi$ on an inflaton field of odd modular weight, and thus provides the $\mathbb{Z}_2$ symmetry required for topological inflation. We find that $\omega\equiv e^{2\pi i/3}$ is the preferred fixed point of the modulus $\tau$, whereas $\tau=i$ does not lead to the minimal $\mathbb{Z}_2$ topological inflation. We analyze how the model parameters affect the inflationary dynamics, and determine the parameter space consistent with the Planck and ACT results. Finally, we discuss reheating, including the three-body decay through supergravity effects and the decay into right-handed neutrinos whose bare Majorana masses are forbidden by modular invariance.

\section{Modular symmetry}
\label{sec:modular}
We consider the modular group $\Gamma\equiv\SL$ which acts on the upper half plane as~\cite{Feruglio:2017spp}
\begin{equation}
\tau\to\gamma\tau=\frac{a\tau+b}{c\tau+d} \,,\quad
\gamma=\begin{pmatrix}a&b\\c&d\end{pmatrix}\, ,
\label{eq:tau-transformation}
\end{equation}
with $a,b,c,d \in \mathbb{Z}$ and $ad-bc=1$. $\Gamma$ is generated by $\gS$ and $\gT$, and we denote its central element by $\Rc\equiv\gS^2$:
\begin{equation}
\gS=\begin{pmatrix}0&1\\-1&0\end{pmatrix}\, ,\quad
\gT=\begin{pmatrix}1&1\\0&1\end{pmatrix}\, ,\quad
\Rc=\gS^2=\begin{pmatrix}-1&0\\0&-1\end{pmatrix},
\end{equation}
with $(\gS\gT)^3=\Id$, $\Rc^2=\Id$ and $\Rc\gT=\gT\Rc$. Note that $\Rc$ leaves $\tau$ invariant. However, it acts non-trivially on fields of odd weight, which is crucial for the $\mathbb{Z}_2$ symmetry to appear as we shall discuss later. 

In $\mathcal{N}=1$ supergravity the modulus $\tau$ is the scalar component of a chiral superfield:
\begin{equation}
\mathcal T(\theta)=\tau+\sqrt2 \theta\psi_\tau+\theta\theta F^\tau \, ,
\label{eq:tau-superfield}
\end{equation}
where $\theta$ is the Grassmann coordinate of superspace, and $\psi_\tau$ and $F^\tau$ denote the modulino field and auxiliary field, respectively. $\mathcal T$ obeys the transformation law in Eq.~\eqref{eq:tau-transformation}, with $\tau$ replaced by $\mathcal T$. Matter chiral superfields $\Phi_I$ carry modular weights $k_I$ and transform as
\begin{equation}
\Phi_I\to(c\mathcal T+d)^{-k_I}\rho_I(\gamma)\,\Phi_I \, .
\label{eq:modtrans}
\end{equation}
In this letter all fields are in the trivial representation of level one, $\rho_I=1$. 

A modular form evaluated on $\mathcal T$ defines a composite chiral superfield, $Y(\mathcal T)=Y(\tau)+\sqrt2\,\theta\psi_\tau Y'(\tau)+\theta\theta[F^\tau Y'(\tau)-\psi_\tau\psi_\tau Y''(\tau)/2]$, where $Y(\tau)$ of weight $k$ and level one is a holomorphic function satisfying $Y(\gamma\tau)=(c\tau+d)^kY(\tau)$, and primes denote derivatives with respect to $\tau$. Under the $\gamma=\Rc$ transformation, $Y=(-1)^kY$, so nonvanishing modular forms have even weight. The ring of level-one modular forms is generated by the Eisenstein series
\begin{align}
E_4(\tau)&=1+240\sum_{n\ge1}\sigma_3(n)q^n\, , \label{eq:E4} \\
E_6(\tau)&=1-504\sum_{n\ge1}\sigma_5(n)q^n\,, \label{eq:E6}
\end{align}
where $q\equiv e^{2\pi i\tau}$ and $\sigma_k(n)=\sum_{d|n}d^k$. 

As points related by a modular transformation describe physically equivalent vacua, it is sufficient to study the dynamics of $\tau$ within the fundamental domain~\cite{Novichkov:2018ovf}, which contains two elliptic fixed points, $\tau=i$ and $\tau=\omega$. The former is invariant under $\gS$, while the latter is invariant under $\gS\gT$. Including $\Rc$, the residual symmetries of $\SL$ at $\tau=i$ and $\tau=\omega$ are $\mathbb{Z}_4^{\gS}$ and $\mathbb{Z}_3^{\gS\gT}\times\mathbb{Z}_2^{\Rc}$, respectively.

\section{The topological inflation model}
\label{sec:model}

We take the modulus K\"ahler potential to be $K\supset-h\ln\mathcal S$, with $\mathcal S \equiv -i(\mathcal T-\overline{\mathcal T})$ and $h$ being a positive integer measuring the number of complex dimensions controlled by ${\mathcal T}$. Under the modular transformation, $K$ undergoes a K\"ahler transformation $K\to K+F+\bar F$, with $F(\mathcal T)=h\ln(c\mathcal T+d)$~\cite{Ferrara:1989bc}. Accordingly, the superpotential must transform as $W\to (c\mathcal T+d)^{-h}W$ to keep the supergravity action modular invariant. Note that we work in reduced Planck units $\Mp=1$. 

In the inflaton sector, we introduce two chiral superfields, the inflaton $\phi$ and the stabilizer $X$, as in Refs.~\cite{Izawa:1998rh,Kawasaki:2000tv}. Apart from $\SL$, we also introduce a $\Ur$ symmetry with $R[\theta]=1$ and $R[W]=2$ to forbid a constant in $W$ and guarantee an $R$-invariant Minkowski vacuum~\cite{Izawa:1998rh}. The inflaton $\phi$ is assigned weight $k_\phi=3$ and $R[\phi]=0$, and $X$ is assigned weight $k_X=h$ and $R[X]=2$. The choice $k_X=h$ makes the coefficient of the linear term $X$ in $W$ a constant. The inflaton
$\phi$ should have an odd weight, so that it is odd under $\Rc$, and $k_\phi=3$ is the smallest choice allowing an $X\phi^2$ term, as there is no level-one modular form of weight two.

The central element $\Rc$ leaves $\tau$ invariant and hence it is not broken by the modulus vacuum. It acts as $\phi\to-\phi$, $X\to(-1)^hX$ and $W\to(-1)^hW$. For even $h$ this is exactly the $\mathbb{Z}_2$ of Refs.~\cite{Izawa:1998rh,Kawasaki:2000tv}, while for odd $h$ it is a $\mathbb{Z}_2^R$, which reduces to the same $\mathbb{Z}_2$ after combining with a $\Ur$ rotation by $\alpha=\pi/2$.\footnote{The fields $\phi$ and $X$, as well as the right-handed neutrinos $N$ introduced in Sec.~\ref{sec:reheating}, are gauge singlets and hence do not induce additional mixed anomalies between this $\mathbb{Z}_2$ and the Standard Model gauge groups. Moreover, a $\mathbb{Z}_2$ symmetry is free from pure gravitational anomalies in four dimensions~\cite{Garcia-Etxebarria:2018ajm}. The anomaly of the full modular group depends on the full spectrum and is beyond the scope of this paper.}

Given the field content listed above, the general K\"ahler potential can be written as
\begin{align}
K=&-h\ln\mathcal S+\frac{|X|^2}{\mathcal S^{h}}+\frac{|\phi|^2}{\mathcal S^{3}} \nonumber \\
&+k_1\frac{|X|^2|\phi|^2}{\mathcal S^{h+3}}+\frac{k_2}{4}\frac{|X|^4}{\mathcal S^{2h}}+\frac{k_3}{4}\frac{|\phi|^4}{\mathcal S^{6}}+\cdots ,
\label{eq:K}
\end{align}
with $k_i$ being real coefficients. The first term is modular invariant up to a K\"ahler transformation, while the remaining terms are precisely invariant under the modular transformation. 

Now we consider the superpotential $W$. The coefficient of $X\phi^{n}$ must be a modular form of weight $k_X+3n-h=3n$. For odd $n$ it vanishes identically, and for $n=0,2$ the spaces of weight-zero and weight-six modular forms are one-dimensional, spanned by $1$ and $E_6$, respectively. At weight twelve, however, there are two independent forms, $E_4^3$ and $E_6^2$. The superpotential is therefore
\begin{align}
W=v^2X&\big[1-\lambda E_6(\mathcal T)\phi^2 \nonumber \\
&+\big(\lambda_1E_6(\mathcal T)^2+\lambda_2E_4(\mathcal T)^3\big)\phi^4+\cdots\big] \, ,
\label{eq:W}
\end{align}
where $E_{4,6}(\mathcal T)$ are the chiral superfields whose scalar components are given in Eqs.~\eqref{eq:E4} and \eqref{eq:E6}, $v$, $\lambda$ and $\lambda_{1,2}$ are constants, and we have omitted terms of higher order than $X\phi^4$.\footnote{The terms $X\phi^{2n}$ with $n\geq2$ are not forbidden by modular symmetry. Nor can they be neglected on the grounds of small field values, since $|\phi|$ is not small during inflation, namely, $|\phi|^2\simeq0.1$--$0.2$ (canonically normalized) when the CMB scales leave the horizon, and $|\phi|=\mathcal O(1)$ at the end of inflation. As we shall see later, the leading higher-order effects come from $X\phi^4$, whereas the other higher-order terms are suppressed relative to it by further powers of $|\phi|^2$. This is why we keep terms in $W$ up to $\mathcal O(\phi^4)$. The same applies to the K\"ahler potential~\eqref{eq:K}, in which we keep terms up to $\mathcal O(|\phi|^4)$.} In general $\lambda_{1,2}$ carry phases that cannot be removed by field redefinitions. However, we can impose a generalized CP symmetry, $\tau\to-\bar\tau$~\cite{Novichkov:2019sqv}, under which all constants in Eq.~\eqref{eq:W} are real, as the Eisenstein series have real Fourier coefficients. Without loss of generality, $v^2$ and $\lambda$ can be taken positive by the redefinitions $X\to-X$ and $\phi\to i\phi$, respectively.

Since the transformation in Eq.~\eqref{eq:modtrans} depends on $\mathcal T$, it also mixes the matter fermions with the modulino $\psi_\tau$. We assume that the modulus $\tau$ is stabilized before inflation, and that the modulus and modulino masses are much larger than both the inflationary Hubble scale and the inflaton mass. Then we can integrate out the heavy modulus multiplet and set $\mathcal T=\tau_0\equiv\langle\tau\rangle$ with $\psi_\tau=0$ in the effective theory. Consequently, this mixing decouples. We further require $\langle F^\tau\rangle=0$ in the stabilizing sector, otherwise $\mathcal T\to\tau_0+\theta\theta F^\tau$ and $Y(\mathcal T)\to Y(\tau_0)+\theta\theta F^\tau Y'(\tau_0)$ would generate soft supersymmetry-breaking terms. Evaluating $K$ and $W$ on the scalar components, we obtain the scalar potential in supergravity:
\begin{equation}
V=e^{K}(K^{i\bar j}D_iW \overline{D_jW}-3|W|^2) \, ,
\label{eq:Vsugra}
\end{equation}
where the indices $i,j$ run over the matter superfields $\Phi_I$, $K_{i\bar j}\equiv\partial_i\partial_{\bar j}K$ is the K\"ahler metric, $K^{i\bar j}$ is its inverse, and $D_iW=W_i+K_iW$ with subscripts denoting derivatives with respect to their scalar components.

From now on we fix $\tau$ specifically at $\tau_0=\omega$ for the following reasons:
\begin{itemize} 
\item  At $\tau=\omega$, $E_6(\omega)\neq0$, and apart from the $X\phi^4$ term $W$ reduces to that of Refs.~\cite{Izawa:1998rh,Kawasaki:2000tv}, with a $\mathbb{Z}_2$-symmetric pair of vacua, see Eq.~\eqref{eq:W-omega} below. At $\tau=i$, instead, $E_6(i)=0$ removes the $X\phi^2$ term, and the leading $\phi$-dependence is $W=v^2X[1+\lambda_2E_4(i)^3\phi^4+\cdots]$. The scalar potential is then $U(1)$-symmetric at $\mathcal{O}(|\phi|^2)$, and the phase dependence first appears at $\mathcal O(\phi^4)$, where $\phi\to-i\phi$ under $\gS$ makes the supersymmetric vacua a $\mathbb{Z}_4$ orbit. This goes beyond the minimal $\mathbb{Z}_2$ topological inflation. In particular, since the phase of $\phi$ is lifted only at $\mathcal O(\phi^4)$, inflation is in general of two-field type.

\item The inflaton potential depends on $\tau$ through the coefficients of $X\phi^{2n}$ in $W$ and the K\"ahler metric of $\phi$, which generically induces a shift $\delta\tau\sim H^2|\phi|^2/m_\tau^2$ of the modulus. Such a shift is negligible for $m_\tau\gg H$ as assumed above, but at $\tau=\omega$ it vanishes exactly: the residual $\mathbb{Z}_3^{\gS\gT}$ leaves $\phi$ invariant, $\phi\to(-\omega-1)^{-3}\phi=\phi$, while $\delta\tau\equiv\tau-\omega\to\omega^2\delta\tau$, so no term linear in $\delta\tau$ appears in $V$ along $X=0$ to all orders in $\phi$. Fixing $\mathcal T=\omega$ is thus self-consistent.

\item Minima of modular-invariant potentials at or near the elliptic fixed points arise in many supergravity models, e.g., from gaugino condensation or with matter contributions~\cite{Font:1990nt,Cvetic:1991qm,Gonzalo:2018guu,Novichkov:2022wvg,Leedom:2022zdm, Knapp-Perez:2023nty,King:2023snq}. In particular, 
$\tau=\omega$ is a natural vacuum for the modulus. Moreover, $\tau=\omega$ is statistically favored in the landscape of flux vacua~\cite{Ishiguro:2020tmo,DeWolfe:2004ns,Ishiguro:2025pwa}.
\end{itemize}

At $\tau=\omega$, one has $-i(\omega-\bar\omega)=\sqrt3$. Hence we rescale $X\to3^{h/4}X$ and $\phi\to3^{3/4}\phi$ so that their kinetic terms are canonical at the origin. The K\"ahler potential in Eq.~\eqref{eq:K} becomes
\begin{equation}
K=|X|^2+|\phi|^2+k_1|X|^2|\phi|^2+\frac{k_2}{4}|X|^4+\frac{k_3}{4}|\phi|^4+\cdots \, ,
\label{eq:K-omega}
\end{equation}
and the superpotential in Eq.~\eqref{eq:W} is written as
\begin{equation}
W=v^2X(1-g\phi^2+g_4\phi^4) \, ,
\label{eq:W-omega}
\end{equation}
where $g\equiv3^{3/2}E_6(\omega)\lambda\simeq15\lambda$, and $g_4\equiv27E_6(\omega)^2\lambda_1$. Here the $\lambda_2$ term drops out because $E_4(\omega)=0$. The overall factor $3^{h/4}$ in $W$ produced by the rescaling of $X$ is removed together with the constant $-h\ln\sqrt3$ in $K$ by a K\"ahler transformation. For $g_4=0$, Eqs.~\eqref{eq:K-omega} and \eqref{eq:W-omega} are precisely the K\"ahler potential and superpotential of the topological inflation model in Ref.~\cite{Kawasaki:2000tv}. The $F$-term conditions $D_XW=D_\phi W=0$ then lead to the two degenerate supersymmetric vacua $\phi_0^2 \equiv \langle\phi\rangle^2=1/g$ related by $\Rc$, together with $\langle X\rangle=0$. For $g_4\neq0$ the pair of vacua $\phi^2_0=1/g$ is shifted to $1/g+g_4/g^3$.

After the rescaling of $X$ and $\phi$, the Lagrangian of the inflaton sector at $\tau=\omega$ and $X=0$ reads
\begin{equation}
\mathcal{L}=(1+k_3|\phi|^2)|\partial\phi|^2 - V \, ,
\end{equation}
with
\begin{equation}
V=v^4\frac{|1-g\phi^2+g_4\phi^4|^2}{1+k_1|\phi|^2}e^{|\phi|^2+\frac{k_3}{4}|\phi|^4} \, .
\end{equation}
We decompose $\phi=(\varphi+i\chi)/\sqrt2$, and the scalar potential along the real direction $\chi=0$ becomes
\begin{equation}
V(\varphi)=v^4\left(1-\frac g2\varphi^2+\frac{g_4}{4}\varphi^4\right)^2\frac{\exp\left(\frac12\varphi^2+\frac{k_3}{16}\varphi^4\right)}{1+\frac{k_1}{2}\varphi^2} \, .
\label{eq:Vinf}
\end{equation}
For $g_4=0$, the above equation is precisely Eq.~(30) of Ref.~\cite{Kawasaki:2000tv}.

\section{Inflationary dynamics}
\label{sec:inflation}
In terms of the canonical field $\hat\varphi$, $d\hat\varphi=\sqrt f\,d\varphi$ with $f=1+k_3\varphi^2/2$, the slow-roll parameters are
\begin{equation}
\epsilon_V=\frac{1}{2f}\left(\frac{V'}{V}\right)^2 \, ,\quad
\eta_V=\frac1f\left(\frac{V''}{V}-\frac{f'}{2f}\frac{V'}{V}\right) \, ,
\label{eq:slowroll}
\end{equation}
where primes denote derivatives with respect to $\varphi$. In the slow-roll approximation, the spectral index $n_s$, the tensor-to-scalar ratio $r$ and the scalar amplitude $A_s$ are given by
\begin{equation}
n_s\simeq1-6\epsilon_V+2\eta_V \, ,\quad r\simeq16\epsilon_V\, ,\quad A_s\simeq\frac{V}{24\pi^2\epsilon_V} \, ,
\label{eq:obs}
\end{equation}
evaluated at $\varphi_*$, where the pivot scale $k_*=0.05\,{\rm Mpc}^{-1}$ leaves the horizon. The number of e-folds from $\varphi_*$ to the end of inflation is
\begin{equation}
N_*\simeq\int_{\varphi_*}^{\varphi_{\rm end}}\frac{f V}{|V'|}d\varphi \, ,
\label{eq:Nefold}
\end{equation}
where $\varphi_{\rm end}$ is determined by $\epsilon_V(\varphi_{\rm end})\simeq1$.

\begin{figure*}[t!]
\centering
\includegraphics[width=\textwidth]{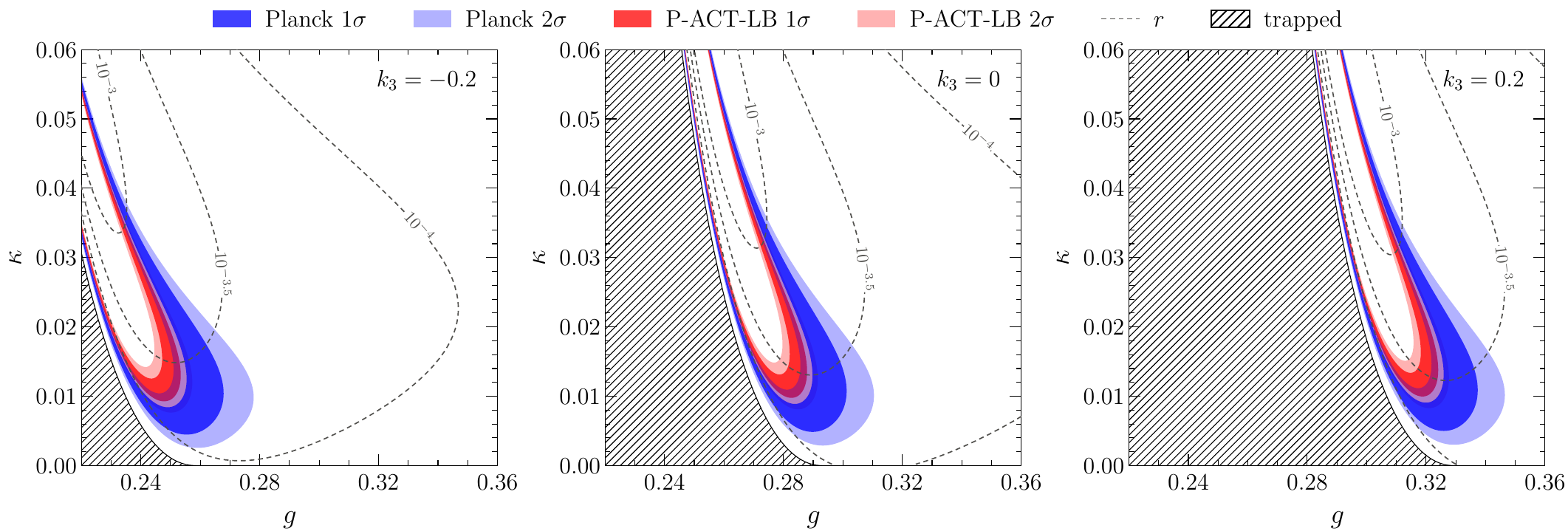}
\caption{Constraints from the spectral index $n_s$ on the potential~\eqref{eq:Vinf} in the $(g,\kappa)$ plane, where $\kappa=2g+k_1-1$. The left, middle and right panels correspond to $k_3=-0.2$, $0$ and $0.2$, respectively, with $g_4=0$ and $N_*\simeq49$--$52$. The dark (light) blue regions denote the $1\sigma$ ($2\sigma$) allowed ranges of $n_s$ from Planck 2018, while the dark (light) red regions denote those from P-ACT-LB. The dashed lines are contours of the tensor-to-scalar ratio $r$, labeled by their values. In the hatched regions, the inflaton is trapped in a local minimum before reaching the vacuum.}
\label{fig:ns}
\end{figure*}

The Planck 2018 data combined with BICEP/Keck 2018 give~\cite{Planck:2018jri,BICEP:2021xfz}
\begin{equation}
\begin{aligned}
&n_s=0.9649\pm0.0042 &&(68\%\ {\rm C.L.}) \, ,\\
&r<0.036 &&(95\%\ {\rm C.L.}) \, .
\end{aligned}
\label{eq:Planck}
\end{equation}
More recently, ACT DR6 combined with Planck, CMB lensing and DESI DR1 baryon acoustic oscillations, together with BICEP/Keck 2018 for the tensor modes, favors a larger spectral index~\cite{AtacamaCosmologyTelescope:2025blo,AtacamaCosmologyTelescope:2025nti},
\begin{equation}
\begin{aligned}
&n_s=0.9743\pm0.0034 &&(68\%\ {\rm C.L.,\ P\text{-}ACT\text{-}LB}) \, ,\\
&r<0.038 &&(95\%\ {\rm C.L.,\ P\text{-}ACT\text{-}LB\text{-}BK18}) \, .
\end{aligned}
\label{eq:ACT}
\end{equation}
Both measurements are obtained at $k_*=0.05\,{\rm Mpc}^{-1}$, where the best-fit amplitude of the scalar perturbations is $A_s\simeq2.1\times10^{-9}$.

Although the inflationary observables will be computed numerically with the full potential~\eqref{eq:Vinf}, the role of the different orders in the inflationary dynamics can be quantified by expanding the potential at small $\varphi$,
\begin{equation}
\frac{V}{v^4}=1-\frac{\kappa}{2}\varphi^2+\alpha_4\varphi^4+\alpha_6\varphi^6+\mathcal O(\varphi^8) \, ,
\label{eq:hilltop}
\end{equation}
with
\begin{align}
\kappa&\equiv2g+k_1-1 \, ,\nonumber\\
\alpha_4&\equiv\frac{\kappa^2+k_1^2-2g^2}{8}+\frac{k_3}{16}+\frac{g_4}{2} \, ,\nonumber\\
\alpha_6&\equiv\frac{g(1-g)^2}{4}-\frac1{24}-\frac{\kappa(10g^2-12g+3)}{16}\nonumber\\
&\quad+\frac{\kappa^2(2g-1)}{4}-\frac{\kappa^3}{8}-\frac{k_3\kappa}{32}+\frac{g_4(g-\kappa)}{4} \, .
\label{eq:coeffs}
\end{align}
With the quadratic term alone, $N_*\simeq\kappa^{-1}\ln(\varphi_{\rm end}/\varphi_*)$ and $n_s\simeq1-2\kappa$~\cite{Izawa:1998rh,Kawasaki:2000tv}, so that $n_s\simeq0.97$ requires $\kappa=\mathcal O(10^{-2})$. Then the condition $\epsilon_V\simeq1$ would be reached only at $\varphi_{\rm end}\simeq\sqrt{2/\kappa}=\mathcal O(10)$, far beyond the vacuum at $\varphi_0\simeq\sqrt{2/g}=\mathcal O(1)$. Inflation must therefore be ended by the higher-order terms, whose contributions to the inflationary observables cannot be neglected. 

The relevant model parameters are $v$, $g$, $k_1$, $k_3$ and $g_4$, while $k_2$ does not enter along the $\varphi$ direction and only needs to satisfy $k_2\lesssim-1$ to stabilize $X$ at the origin during inflation~\cite{Kawasaki:2000tv}. The overall scale $v$ is fixed by $A_s$ for given $N_*$. As the predicted $r$ lies far below the observational bound, the remaining four parameters are constrained essentially by $n_s$ alone. The parameters $g$ and $k_1$ enter the coefficients at all orders. In contrast, $k_3$ and $g_4$ enter $\alpha_4$ only through the combination $k_3/16+g_4/2$, while their contributions to $\alpha_6$ are suppressed. Hence the observables depend on them approximately only through this combination, and they mainly shift the viable parameter space of $g$ and $\kappa$, leaving its shape almost unchanged. In the following we set $g_4=0$ and vary only $k_3$, which effectively accounts for both couplings.

Fig.~\ref{fig:ns} shows the $1\sigma$ and $2\sigma$ regions consistent with the Planck and P-ACT-LB results. The allowed parameter space is U-shaped, while the region enclosed by it is excluded due to large $n_s$ there. As expected, $k_3$ mainly shifts the viable region along $g$ while its shape is almost unchanged. For $k_3=0$, the Planck constraint yields $g\simeq0.25$--$0.31$ and $k_1\simeq0.39$--$0.57$ at the $2\sigma$ level, consistent with Ref.~\cite{Choi:2022fce}. At smaller $g$, $\alpha_4$ increases. Once $\alpha_4>0$, $\alpha_6<0$ and $\alpha_4^2>3\kappa|\alpha_6|/2$, $V'$ has a pair of zeros between the origin and the vacuum, so that the inflaton starting from the core of the wall is trapped in a local minimum. The P-ACT-LB data shift the allowed parameter space toward the inner side of the U-shaped region and make it smaller, due to the smaller uncertainty and to the larger central value of $n_s$, around which $n_s$ depends more sensitively on the model parameters. The $1\sigma$ allowed regions for Planck and P-ACT-LB do not overlap. Nevertheless, there is a common parameter space consistent with both data sets at the $2\sigma$ level, corresponding to $0.9675\lesssim n_s\lesssim0.9733$.

\section{Reheating}
\label{sec:reheating}
After inflation the universe is reheated by the decay of the inflaton oscillating around one of the vacua $\phi_0$, where we take $\phi_0>0$ without loss of generality. In the parameter region compatible with the data, we have $\phi_0\simeq1.9$ and the inflaton mass is $m_\phi\simeq3\times10^{13}\,$GeV. The reheating temperature is given by $T_R=(\pi^2g_*/10)^{-1/4}\sqrt{\Gamma_\phi}$, with $\Gamma_\phi$ the total decay rate and $g_*=228.75$ the number of relativistic degrees of freedom. 

One possible decay channel of $\phi$ is into right-handed neutrinos $N$~\cite{Kawasaki:2000tv,Choi:2022fce,Kawasaki:2023mjm}, whose couplings to $\phi$ are constrained by the modular covariance of the superpotential. We take $R[N]=1$. For odd $h$, the coupling $y_N\phi NN/2$ with a constant $y_N$ is allowed in the superpotential if we assign the weight $k_N=(h-3)/2$ to the right-handed neutrinos. This coupling generates the Majorana mass $M_N=y_N\phi_0$ in the vacuum. For even $h$, $\phi NN$ is forbidden by $\Rc$. With $k_N=(h-6)/2$, the leading coupling becomes $y_N\phi^2NN/2$, which leads to $M_N=y_N\phi_0^2$. In both cases a bare Majorana mass would require a coefficient of negative weight and is therefore forbidden. For odd $h$, the total decay rate into right-handed neutrinos and sneutrinos, $\phi\to NN$ and $\phi\to\tilde N\tilde N^*$, is
\begin{equation}
\Gamma_N\simeq\frac{|y_N|^2}{32\pi}m_\phi\left(1-\frac{4M_N^2}{m_\phi^2}\right)^{1/2} \, .
\label{eq:GammaN}
\end{equation}
Kinematics requires $M_N<m_\phi/2$, i.e.\ $y_N<m_\phi/(2\phi_0)\simeq3\times10^{-6}$, and $\Gamma_N$ is maximal at $M_N=m_\phi/\sqrt6\simeq10^{13}\,$GeV, where this channel alone gives $T_R\simeq5\times10^8\,$GeV. For even $h$ the effective coupling $2y_N\phi_0$ allows $T_R\lesssim9\times10^8\,$GeV. When $T_R\ll M_N$, the right-handed (s)neutrinos decay out of equilibrium and can generate the baryon asymmetry through non-thermal leptogenesis~\cite{Fukugita:1986hr,Asaka:1999yd,Asaka:1999jb,Kawasaki:2023mjm}.

Moreover, since $\phi_0\neq0$, the inflaton couples to all fields in $W$ through supergravity effects~\cite{Endo:2006qk,Endo:2007sz}. The dominant contribution is the three-body decay via the top Yukawa coupling $y_t$. Assuming the minimal K\"ahler potential for the matter fields, this decay channel is described by the effective superpotential term $W_{\rm eff}\supset y_t\phi_0\delta\phi H_uQ\bar U$ with $\delta\phi\equiv\phi-\phi_0$, and the decay rate is~\cite{Endo:2007sz}
\begin{equation}
\Gamma_t\simeq\frac{3}{256\pi^3}|y_t\phi_0|^2 m_\phi^3 \, ,
\label{eq:Gammat}
\end{equation}
which alone gives $T_R\simeq6\times10^8\,{\rm GeV}$ for $y_t\simeq0.6$. Since $\Gamma_N\propto m_\phi^3/\phi_0^2$ and $\Gamma_t\propto\phi_0^2m_\phi^3$, the ratio $\Gamma_N/\Gamma_t\lesssim0.6$ for odd $h$ and $\lesssim2.3$ for even $h$ is independent of $v$. Hence the two channels are comparable, and in total $T_R\simeq10^9\,$GeV. The inflaton oscillation behaves as matter until $H\simeq\Gamma_\phi$, so the number of e-folds $N_*$ depends on the reheating temperature only through $(\ln T_R)/3$~\cite{Choi:2022fce}. This gives $N_*\simeq49$--$52$ over the parameter region of interest.

Through the same supergravity effects the inflaton also decays into gravitinos, both directly and via the supersymmetry-breaking sector~\cite{Endo:2006zj,Nakamura:2006uc,Kawasaki:2006gs,Endo:2007sz}. For $\phi_0\simeq1.9$, $m_\phi\simeq3\times10^{13}\,$GeV and $T_R\simeq10^9\,$GeV, the resulting gravitino abundance is $Y_{3/2}\sim10^{-7}$--$10^{-3}$, depending on the supersymmetry-breaking sector~\cite{Endo:2007sz}. Such a large abundance exceeds the bounds from big-bang nucleosynthesis and the dark matter abundance by many orders of magnitude. This constraint can be easily avoided, e.g., for a gravitino lighter than $\mathcal O(10)\,$eV~\cite{Endo:2007sz}. 

\section{Conclusions}
\label{sec:conclusions}
We have shown that the topological inflation model of Refs.~\cite{Izawa:1998rh,Kawasaki:2000tv,Choi:2022fce,Choi:2022ssv} can be embedded in a theory with $\SL$ modular symmetry. The inflaton $\phi$ is assigned the odd modular weight $k_\phi=3$, so that $\Rc$ acts as $\phi\to-\phi$ and provides the $\mathbb{Z}_2$ symmetry required for topological inflation. We find that $\tau=\omega$ is a natural choice for the modulus stabilization in our scenario. It reproduces the $\mathbb{Z}_2$-symmetric pair of vacua required by the minimal topological inflation model, and is favored by many supergravity models of modulus stabilization and the flux landscape. In contrast, at $\tau=i$ modular symmetry requires $E_6(i)=0$, which removes the $X\phi^2$ term. As a result, the minimal $\mathbb{Z}_2$ topological inflation is not realized. Modular symmetry does not forbid higher-order terms in the inflaton field. Since $|\phi|$ is not small during inflation, their contributions should be taken into account. The leading ones are the quartic couplings $k_3$ in the K\"ahler potential and $g_4$ in the superpotential, which mainly shift the viable parameter space. We have solved the inflationary dynamics numerically and constrained the parameter space with the Planck and P-ACT-LB results. The viable region for P-ACT-LB is narrower, due to the smaller uncertainty and to the larger central value of $n_s$, around which $n_s$ depends more sensitively on the model parameters. We have also studied reheating after inflation. When right-handed neutrinos $N$ are introduced, modular symmetry forbids their bare Majorana masses while allowing the coupling $\phi NN$ for odd $h$ (or $\phi^2NN$ for even $h$). This coupling opens the decay channel $\phi\to NN$, which allows non-thermal leptogenesis. Including also the supergravity-induced three-body decay into the top sector, the reheating temperature is $T_R\simeq10^9\,$GeV. The inflaton also decays into gravitinos, and the resulting abundance is compatible with the constraints from big-bang nucleosynthesis and the dark matter abundance if, e.g., the gravitino is lighter than $\mathcal O(10)\,$eV.

Our results suggest that modular symmetry may play an important role in the inflationary epoch of the universe. Possible extensions include more general $\mathbb{Z}_N$ topological inflation, e.g., based on the residual $\mathbb{Z}_4$ symmetry at $\tau=i$, and multi-field inflation with several inflatons, e.g., transforming as nontrivial representations of the finite modular groups at higher levels. In addition, the connection between the inflationary sector and the low-energy flavor sector deserves further study.

\begin{acknowledgments}
T.~T.~Y. thanks Serguey~T.~Petcov for serious discussions on modular symmetry. X.~W.~was supported by the European Union--NextGenerationEU
through the National Recovery and Resilience Plan
(Mission~4, Component~2),
under the project
\textit{MODIPAC: Modular Invariance in Particle Physics and Cosmology}
(CUP~C93C24004940006).
X.~W.'s research visit to Kavli IPMU
was supported by the ASYMMETRY project,
funded by the European Union's Horizon Europe programme
under the Marie Sk{\l}odowska-Curie Actions Staff Exchanges
(Grant Agreement No.~101086085). T.~T.~Y. is supported by MEXT Grant No.~24H02244 and the World Premier International Research Center Initiative (WPI), MEXT, Japan (Kavli IPMU).

\end{acknowledgments}

\bibliographystyle{name}
\bibliography{references}

\providecommand{\href}[2]{#2}\begingroup\raggedright\begin{thebibliography}{10}

\bibitem{Starobinsky:1980te}
A.~A.~Starobinsky, {\em {A New Type of Isotropic Cosmological Models Without Singularity}}, \href{https://dx.doi.org/10.1016/0370-2693(80)90670-X}{Phys.\  Lett.\  B {\bfseries 91} (1980) 99--102}.

\bibitem{Guth:1980zm}
A.~H.~Guth, {\em {The Inflationary Universe: A Possible Solution to the Horizon and Flatness Problems}}, \href{https://dx.doi.org/10.1103/PhysRevD.23.347}{Phys.\  Rev.\  D {\bfseries 23} (1981) 347--356}.

\bibitem{Linde:1981mu}
A.~D.~Linde, {\em {A New Inflationary Universe Scenario: A Possible Solution of the Horizon, Flatness, Homogeneity, Isotropy and Primordial Monopole Problems}}, \href{https://dx.doi.org/10.1016/0370-2693(82)91219-9}{Phys.\  Lett.\  B {\bfseries 108} (1982) 389--393}.

\bibitem{Albrecht:1982wi}
A.~Albrecht and P.~J.~Steinhardt, {\em {Cosmology for Grand Unified Theories with Radiatively Induced Symmetry Breaking}}, \href{https://dx.doi.org/10.1103/PhysRevLett.48.1220}{Phys.\  Rev.\  Lett.\  {\bfseries 48} (1982) 1220--1223}.

\bibitem{Planck:2018jri}
{\bfseries Planck} Collaboration, {\em {Planck 2018 results. X. Constraints on inflation}}, \href{https://dx.doi.org/10.1051/0004-6361/201833887}{Astron.\  Astrophys.\  {\bfseries 641} (2020) A10} {\ttfamily [\href{https://arxiv.org/abs/1807.06211}{arXiv:1807.06211}]}.

\bibitem{BICEP:2021xfz}
{\bfseries BICEP, Keck} Collaboration, {\em {Improved Constraints on Primordial Gravitational Waves using Planck, WMAP, and BICEP/Keck Observations through the 2018 Observing Season}}, \href{https://dx.doi.org/10.1103/PhysRevLett.127.151301}{Phys.\  Rev.\  Lett.\  {\bfseries 127} (2021) 151301} {\ttfamily [\href{https://arxiv.org/abs/2110.00483}{arXiv:2110.00483}]}.

\bibitem{AtacamaCosmologyTelescope:2025blo}
{\bfseries Atacama Cosmology Telescope} Collaboration, {\em {The Atacama Cosmology Telescope: DR6 power spectra, likelihoods and {\ensuremath{\Lambda}}CDM parameters}}, \href{https://dx.doi.org/10.1088/1475-7516/2025/11/062}{JCAP {\bfseries 11} (2025) 062} {\ttfamily [\href{https://arxiv.org/abs/2503.14452}{arXiv:2503.14452}]}.

\bibitem{Izawa:1998rh}
K.~I.~Izawa, M.~Kawasaki, and T.~Yanagida, {\em {R invariant topological inflation}}, \href{https://dx.doi.org/10.1143/PTP.101.1129}{Prog.\  Theor.\  Phys.\  {\bfseries 101} (1999) 1129--1133} {\ttfamily [\href{https://arxiv.org/abs/hep-ph/9810537}{hep-ph/9810537}]}.

\bibitem{Antusch:2014qqa}
S.~Antusch, D.~Nolde, and S.~Orani, {\em {Hilltop inflation with preinflation from coupling to matter fields}}, \href{https://dx.doi.org/10.1088/1475-7516/2014/05/034}{JCAP {\bfseries 05} (2014) 034} {\ttfamily [\href{https://arxiv.org/abs/1402.5328}{arXiv:1402.5328}]}.

\bibitem{Ai:2025qua}
W.-Y.~Ai, S.~F.~King, X.~Wang, and Y.-L.~Zhou, {\em {Curvaton-assisted hilltop inflation}}, \href{https://dx.doi.org/10.1103/gbgy-7fkl}{Phys.\  Rev.\  D {\bfseries 114} (2026) 076002} {\ttfamily [\href{https://arxiv.org/abs/2512.09079}{arXiv:2512.09079}]}.

\bibitem{Linde:1994hy}
A.~D.~Linde, {\em {Monopoles as big as a universe}}, \href{https://dx.doi.org/10.1016/0370-2693(94)90719-6}{Phys.\  Lett.\  B {\bfseries 327} (1994) 208--213} {\ttfamily [\href{https://arxiv.org/abs/astro-ph/9402031}{astro-ph/9402031}]}.

\bibitem{Vilenkin:1994pv}
A.~Vilenkin, {\em {Topological inflation}}, \href{https://dx.doi.org/10.1103/PhysRevLett.72.3137}{Phys.\  Rev.\  Lett.\  {\bfseries 72} (1994) 3137--3140} {\ttfamily [\href{https://arxiv.org/abs/hep-th/9402085}{hep-th/9402085}]}.

\bibitem{Linde:1994wt}
A.~D.~Linde and D.~A.~Linde, {\em {Topological defects as seeds for eternal inflation}}, \href{https://dx.doi.org/10.1103/PhysRevD.50.2456}{Phys.\  Rev.\  D {\bfseries 50} (1994) 2456--2468} {\ttfamily [\href{https://arxiv.org/abs/hep-th/9402115}{hep-th/9402115}]}.

\bibitem{Kawasaki:2000tv}
M.~Kawasaki, N.~Sakai, M.~Yamaguchi, and T.~Yanagida, {\em {Topological inflation in supergravity}}, \href{https://dx.doi.org/10.1103/PhysRevD.62.123507}{Phys.\  Rev.\  D {\bfseries 62} (2000) 123507} {\ttfamily [\href{https://arxiv.org/abs/hep-ph/0005073}{hep-ph/0005073}]}.

\bibitem{Choi:2022fce}
G.~Choi, W.~Lin, and T.~T.~Yanagida, {\em {Discrete R-symmetry, various energy scales, and gravitational waves}}, \href{https://dx.doi.org/10.1103/PhysRevD.105.055033}{Phys.\  Rev.\  D {\bfseries 105} (2022) 055033} {\ttfamily [\href{https://arxiv.org/abs/2201.03083}{arXiv:2201.03083}]}.

\bibitem{Choi:2022ssv}
G.~Choi, W.~Lin, and T.~T.~Yanagida, {\em {Dynamically induced topological inflation}}, \href{https://dx.doi.org/10.1103/PhysRevD.107.036005}{Phys.\  Rev.\  D {\bfseries 107} (2023) 036005} {\ttfamily [\href{https://arxiv.org/abs/2204.03675}{arXiv:2204.03675}]}.

\bibitem{Ferrara:1989bc}
S.~Ferrara, D.~Lust, A.~D.~Shapere, and S.~Theisen, {\em {Modular Invariance in Supersymmetric Field Theories}}, \href{https://dx.doi.org/10.1016/0370-2693(89)90583-2}{Phys.\  Lett.\  B {\bfseries 225} (1989) 363}.

\bibitem{Ferrara:1989qb}
S.~Ferrara, D.~L{\"u}st, and S.~Theisen, {\em {Target Space Modular Invariance and Low-Energy Couplings in Orbifold Compactifications}}, \href{https://dx.doi.org/10.1016/0370-2693(89)90631-X}{Phys.\  Lett.\  B {\bfseries 233} (1989) 147--152}.

\bibitem{Feruglio:2017spp}
F.~Feruglio, \href{https://dx.doi.org/10.1142/9789813238053_0012}{{\em {Are neutrino masses modular forms?}},} in {\em {From My Vast Repertoire ...}: {Guido Altarelli's Legacy}}, A.~Levy, {\em et al.}, eds., pp.~227--266.
\newblock World Scientific, 2019.
\newblock {\ttfamily \href{https://arxiv.org/abs/1706.08749}{arXiv:1706.08749}}.

\bibitem{Feruglio:2019ybq}
F.~Feruglio and A.~Romanino, {\em {Lepton flavor symmetries}}, \href{https://dx.doi.org/10.1103/RevModPhys.93.015007}{Rev.\  Mod.\  Phys.\  {\bfseries 93} (2021) 015007} {\ttfamily [\href{https://arxiv.org/abs/1912.06028}{arXiv:1912.06028}]}.

\bibitem{Kobayashi:2023zzc}
T.~Kobayashi and M.~Tanimoto, {\em {Modular flavor symmetric models}}, \href{https://dx.doi.org/10.1142/S0217751X24410124}{Int.\  J.\  Mod.\  Phys.\  A {\bfseries 39} (2024) 2441012} {\ttfamily [\href{https://arxiv.org/abs/2307.03384}{arXiv:2307.03384}]}.

\bibitem{Ding:2023htn}
G.-J.~Ding and S.~F.~King, {\em {Neutrino mass and mixing with modular symmetry}}, \href{https://dx.doi.org/10.1088/1361-6633/ad52a3}{Rept.\  Prog.\  Phys.\  {\bfseries 87} (2024) 084201} {\ttfamily [\href{https://arxiv.org/abs/2311.09282}{arXiv:2311.09282}]}.

\bibitem{Bento:1996xh}
M.~C.~Bento and O.~Bertolami, {\em {Topological inflation in dual superstring models}}, \href{https://dx.doi.org/10.1016/0370-2693(96)00767-8}{Phys.\  Lett.\  B {\bfseries 384} (1996) 98--102} {\ttfamily [\href{https://arxiv.org/abs/gr-qc/9605070}{gr-qc/9605070}]}.

\bibitem{Kobayashi:2016mzg}
T.~Kobayashi, D.~Nitta, and Y.~Urakawa, {\em {Modular invariant inflation}}, \href{https://dx.doi.org/10.1088/1475-7516/2016/08/014}{JCAP {\bfseries 08} (2016) 014} {\ttfamily [\href{https://arxiv.org/abs/1604.02995}{arXiv:1604.02995}]}.

\bibitem{Schimmrigk:2016bde}
R.~Schimmrigk, {\em {Modular Inflation Observables and $j$-Inflation Phenomenology}}, \href{https://dx.doi.org/10.1007/JHEP09(2017)043}{JHEP {\bfseries 09} (2017) 043} {\ttfamily [\href{https://arxiv.org/abs/1612.09559}{arXiv:1612.09559}]}.

\bibitem{Schimmrigk:2017jwa}
R.~Schimmrigk, {\em {Multifield Reheating after Modular $j$-Inflation}}, \href{https://dx.doi.org/10.1016/j.physletb.2018.04.065}{Phys.\  Lett.\  B {\bfseries 782} (2018) 193--197} {\ttfamily [\href{https://arxiv.org/abs/1712.09961}{arXiv:1712.09961}]}.

\bibitem{Lynker:2019joa}
M.~Lynker and R.~Schimmrigk, {\em {Modular Inflation at Higher Level $N$}}, \href{https://dx.doi.org/10.1088/1475-7516/2019/06/036}{JCAP {\bfseries 06} (2019) 036} {\ttfamily [\href{https://arxiv.org/abs/1902.04625}{arXiv:1902.04625}]}.

\bibitem{Abe:2023ylh}
Y.~Abe, T.~Higaki, F.~Kaneko, T.~Kobayashi, and H.~Otsuka, {\em {Moduli inflation from modular flavor symmetries}}, \href{https://dx.doi.org/10.1007/JHEP06(2023)187}{JHEP {\bfseries 06} (2023) 187} {\ttfamily [\href{https://arxiv.org/abs/2303.02947}{arXiv:2303.02947}]}.

\bibitem{King:2024ssx}
S.~F.~King and X.~Wang, {\em {Modular invariant hilltop inflation}}, \href{https://dx.doi.org/10.1088/1475-7516/2024/07/073}{JCAP {\bfseries 07} (2024) 073} {\ttfamily [\href{https://arxiv.org/abs/2405.08924}{arXiv:2405.08924}]}.

\bibitem{Ding:2024neh}
G.-J.~Ding, S.-Y.~Jiang, and W.~Zhao, {\em {Modular invariant slow roll inflation}}, \href{https://dx.doi.org/10.1088/1475-7516/2024/10/016}{JCAP {\bfseries 10} (2024) 016} {\ttfamily [\href{https://arxiv.org/abs/2405.06497}{arXiv:2405.06497}]}.

\bibitem{Casas:2024jbw}
G.~F.~Casas and L.~E.~Ib{\'a}{\~n}ez, {\em {Modular invariant Starobinsky inflation and the Species Scale}}, \href{https://dx.doi.org/10.1007/JHEP04(2025)041}{JHEP {\bfseries 04} (2025) 041} {\ttfamily [\href{https://arxiv.org/abs/2407.12081}{arXiv:2407.12081}]}.

\bibitem{Kallosh:2024ymt}
R.~Kallosh and A.~Linde, {\em {SL(2,{\ensuremath{\mathbb{Z}}}) cosmological attractors}}, \href{https://dx.doi.org/10.1088/1475-7516/2025/04/045}{JCAP {\bfseries 04} (2025) 045} {\ttfamily [\href{https://arxiv.org/abs/2408.05203}{arXiv:2408.05203}]}.

\bibitem{Kallosh:2024pat}
R.~Kallosh and A.~Linde, {\em {Landscape of modular cosmology}}, \href{https://dx.doi.org/10.1088/1475-7516/2025/05/037}{JCAP {\bfseries 05} (2025) 037} {\ttfamily [\href{https://arxiv.org/abs/2411.07552}{arXiv:2411.07552}]}.

\bibitem{Aoki:2024ixq}
S.~Aoki and H.~Otsuka, {\em {Inflationary constraints on the moduli-dependent species scale in modular invariant theories}}, \href{https://dx.doi.org/10.1103/k3xy-5r32}{Phys.\  Rev.\  D {\bfseries 112} (2025) 046006} {\ttfamily [\href{https://arxiv.org/abs/2411.08467}{arXiv:2411.08467}]}.

\bibitem{Ding:2024euc}
G.-J.~Ding, S.-Y.~Jiang, Y.~Xu, and W.~Zhao, {\em {Modular invariant inflation and reheating}}, \href{https://dx.doi.org/10.1007/JHEP11(2025)141}{JHEP {\bfseries 11} (2025) 141} {\ttfamily [\href{https://arxiv.org/abs/2411.18603}{arXiv:2411.18603}]}.

\bibitem{Jiang:2025qbi}
S.-Y.~Jiang, W.~Zhao, and G.-J.~Ding, {\em {Sp(4,{\ensuremath{\mathbb{Z}}}) modular inflation}}, \href{https://dx.doi.org/10.1088/1475-7516/2026/05/090}{JCAP {\bfseries 05} (2026) 090} {\ttfamily [\href{https://arxiv.org/abs/2512.21597}{arXiv:2512.21597}]}.

\bibitem{Abe:2026cmt}
Y.~Abe, K.~Goto, T.~Higaki, J.~Kawamura, and T.~Kobayashi, {\em {Finite modular Coleman-Weinberg inflation}}, {\ttfamily \href{https://arxiv.org/abs/2607.00528}{arXiv:2607.00528}} (2026).

\bibitem{Gunji:2022xig}
Y.~Gunji, K.~Ishiwata, and T.~Yoshida, {\em {Subcritical regime of hybrid inflation with modular A$_{4}$ symmetry}}, \href{https://dx.doi.org/10.1007/JHEP11(2022)002}{JHEP {\bfseries 11} (2022) 002} {\ttfamily [\href{https://arxiv.org/abs/2208.10086}{arXiv:2208.10086}]}.

\bibitem{Aoki:2025wld}
S.~Aoki, H.~Otsuka, and R.~Yanagita, {\em {Higgs-modular inflation}}, \href{https://dx.doi.org/10.1103/v4z9-676d}{Phys.\  Rev.\  D {\bfseries 112} (2025) 043505} {\ttfamily [\href{https://arxiv.org/abs/2504.01622}{arXiv:2504.01622}]}.

\bibitem{Cvetic:1991vp}
M.~Cvetic, F.~Quevedo, and S.-J.~Rey, {\em {Stringy domain walls and target space modular invariance}}, \href{https://dx.doi.org/10.1103/PhysRevLett.67.1836}{Phys.\  Rev.\  Lett.\  {\bfseries 67} (1991) 1836--1839}.

\bibitem{Cvetic:1992bf}
M.~Cvetic, S.~Griffies, and S.-J.~Rey, {\em {Static domain walls in N=1 supergravity}}, \href{https://dx.doi.org/10.1016/0550-3213(92)90649-V}{Nucl.\  Phys.\  B {\bfseries 381} (1992) 301--328} {\ttfamily [\href{https://arxiv.org/abs/hep-th/9201007}{hep-th/9201007}]}.

\bibitem{Cvetic:1992cv}
M.~Cvetic and R.~L.~Davis, {\em {Cosmological implications of domain walls due to duality invariant moduli sector of superstring vacua}}, \href{https://dx.doi.org/10.1016/0370-2693(92)91327-6}{Phys.\  Lett.\  B {\bfseries 296} (1992) 316--322} {\ttfamily [\href{https://arxiv.org/abs/hep-th/9205060}{hep-th/9205060}]}.

\bibitem{Cvetic:1996vr}
M.~Cvetic and H.~H.~Soleng, {\em {Supergravity domain walls}}, \href{https://dx.doi.org/10.1016/S0370-1573(96)00035-X}{Phys.\  Rept.\  {\bfseries 282} (1997) 159--223} {\ttfamily [\href{https://arxiv.org/abs/hep-th/9604090}{hep-th/9604090}]}.

\bibitem{King:2024lki}
S.~F.~King, X.~Wang, and Y.-L.~Zhou, {\em {Modular domain walls and gravitational waves}}, \href{https://dx.doi.org/10.1088/1475-7516/2025/07/011}{JCAP {\bfseries 07} (2025) 011} {\ttfamily [\href{https://arxiv.org/abs/2411.04900}{arXiv:2411.04900}]}.

\bibitem{Nomura:2019jxj}
T.~Nomura and H.~Okada, {\em {A modular $A_4$ symmetric model of dark matter and neutrino}}, \href{https://dx.doi.org/10.1016/j.physletb.2019.134799}{Phys.\  Lett.\  B {\bfseries 797} (2019) 134799} {\ttfamily [\href{https://arxiv.org/abs/1904.03937}{arXiv:1904.03937}]}.

\bibitem{Kobayashi:2021ajl}
T.~Kobayashi, H.~Okada, and Y.~Orikasa, {\em {Dark matter stability at fixed points in a modular A4 symmetry}}, \href{https://dx.doi.org/10.1016/j.dark.2022.101080}{Phys.\  Dark Univ.\  {\bfseries 37} (2022) 101080} {\ttfamily [\href{https://arxiv.org/abs/2111.05674}{arXiv:2111.05674}]}.

\bibitem{Baur:2024lcc}
A.~Baur, M.-C.~Chen, V.~Knapp-Perez, and S.~Ramos-Sanchez, {\em {Modular flavored dark matter}}, \href{https://dx.doi.org/10.1007/JHEP12(2024)091}{JHEP {\bfseries 12} (2024) 091} {\ttfamily [\href{https://arxiv.org/abs/2409.02178}{arXiv:2409.02178}]}.

\bibitem{Anchordoqui:2025epz}
L.~A.~Anchordoqui, {\em et al.}, {\em {Bulk/boundary modular quintessence and DESI}}, \href{https://dx.doi.org/10.1007/JHEP09(2025)128}{JHEP {\bfseries 09} (2025) 128} {\ttfamily [\href{https://arxiv.org/abs/2506.02731}{arXiv:2506.02731}]}.

\bibitem{Gordillo-Ruiz:2025xcg}
H.~Gordillo-Ruiz, M.~Hernandez-Segura, I.~Portillo-Castillo, S.~Ramos-Sanchez, and I.~Zavala, {\em {Rolling with modular symmetry: quintessence and de Sitter in heterotic orbifolds}}, \href{https://dx.doi.org/10.1007/JHEP04(2026)211}{JHEP {\bfseries 04} (2026) 211} {\ttfamily [\href{https://arxiv.org/abs/2509.22781}{arXiv:2509.22781}]}.

\bibitem{Novichkov:2018ovf}
P.~P.~Novichkov, J.~T.~Penedo, S.~T.~Petcov, and A.~V.~Titov, {\em {Modular S$_{4}$ models of lepton masses and mixing}}, \href{https://dx.doi.org/10.1007/JHEP04(2019)005}{JHEP {\bfseries 04} (2019) 005} {\ttfamily [\href{https://arxiv.org/abs/1811.04933}{arXiv:1811.04933}]}.

\bibitem{Garcia-Etxebarria:2018ajm}
I.~Garc{\'\i}a-Etxebarria and M.~Montero, {\em {Dai-Freed anomalies in particle physics}}, \href{https://dx.doi.org/10.1007/JHEP08(2019)003}{JHEP {\bfseries 08} (2019) 003} {\ttfamily [\href{https://arxiv.org/abs/1808.00009}{arXiv:1808.00009}]}.

\bibitem{Novichkov:2019sqv}
P.~P.~Novichkov, J.~T.~Penedo, S.~T.~Petcov, and A.~V.~Titov, {\em {Generalised CP Symmetry in Modular-Invariant Models of Flavour}}, \href{https://dx.doi.org/10.1007/JHEP07(2019)165}{JHEP {\bfseries 07} (2019) 165} {\ttfamily [\href{https://arxiv.org/abs/1905.11970}{arXiv:1905.11970}]}.

\bibitem{Font:1990nt}
A.~Font, L.~E.~Ibanez, D.~Lust, and F.~Quevedo, {\em {Supersymmetry Breaking From Duality Invariant Gaugino Condensation}}, \href{https://dx.doi.org/10.1016/0370-2693(90)90665-S}{Phys.\  Lett.\  B {\bfseries 245} (1990) 401--408}.

\bibitem{Cvetic:1991qm}
M.~Cvetic, A.~Font, L.~E.~Ibanez, D.~Lust, and F.~Quevedo, {\em {Target space duality, supersymmetry breaking and the stability of classical string vacua}}, \href{https://dx.doi.org/10.1016/0550-3213(91)90622-5}{Nucl.\  Phys.\  B {\bfseries 361} (1991) 194--232}.

\bibitem{Gonzalo:2018guu}
E.~Gonzalo, L.~E.~Ib{\'a}{\~n}ez, and {\'A}.~M.~Uranga, {\em {Modular symmetries and the swampland conjectures}}, \href{https://dx.doi.org/10.1007/JHEP05(2019)105}{JHEP {\bfseries 05} (2019) 105} {\ttfamily [\href{https://arxiv.org/abs/1812.06520}{arXiv:1812.06520}]}.

\bibitem{Novichkov:2022wvg}
P.~P.~Novichkov, J.~T.~Penedo, and S.~T.~Petcov, {\em {Modular flavour symmetries and modulus stabilisation}}, \href{https://dx.doi.org/10.1007/JHEP03(2022)149}{JHEP {\bfseries 03} (2022) 149} {\ttfamily [\href{https://arxiv.org/abs/2201.02020}{arXiv:2201.02020}]}.

\bibitem{Leedom:2022zdm}
J.~M.~Leedom, N.~Righi, and A.~Westphal, {\em {Heterotic de Sitter beyond modular symmetry}}, \href{https://dx.doi.org/10.1007/JHEP02(2023)209}{JHEP {\bfseries 02} (2023) 209} {\ttfamily [\href{https://arxiv.org/abs/2212.03876}{arXiv:2212.03876}]}.

\bibitem{Knapp-Perez:2023nty}
V.~Knapp-Perez, X.-G.~Liu, H.~P.~Nilles, S.~Ramos-Sanchez, and M.~Ratz, {\em {Matter matters in moduli fixing and modular flavor symmetries}}, \href{https://dx.doi.org/10.1016/j.physletb.2023.138106}{Phys.\  Lett.\  B {\bfseries 844} (2023) 138106} {\ttfamily [\href{https://arxiv.org/abs/2304.14437}{arXiv:2304.14437}]}.

\bibitem{King:2023snq}
S.~F.~King and X.~Wang, {\em {Modulus stabilization in the multiple-modulus framework}}, \href{https://dx.doi.org/10.1103/PhysRevD.110.076026}{Phys.\  Rev.\  D {\bfseries 110} (2024) 076026} {\ttfamily [\href{https://arxiv.org/abs/2310.10369}{arXiv:2310.10369}]}.

\bibitem{Ishiguro:2020tmo}
K.~Ishiguro, T.~Kobayashi, and H.~Otsuka, {\em {Landscape of Modular Symmetric Flavor Models}}, \href{https://dx.doi.org/10.1007/JHEP03(2021)161}{JHEP {\bfseries 03} (2021) 161} {\ttfamily [\href{https://arxiv.org/abs/2011.09154}{arXiv:2011.09154}]}.

\bibitem{DeWolfe:2004ns}
O.~DeWolfe, A.~Giryavets, S.~Kachru, and W.~Taylor, {\em {Enumerating flux vacua with enhanced symmetries}}, \href{https://dx.doi.org/10.1088/1126-6708/2005/02/037}{JHEP {\bfseries 02} (2005) 037} {\ttfamily [\href{https://arxiv.org/abs/hep-th/0411061}{hep-th/0411061}]}.

\bibitem{Ishiguro:2025pwa}
K.~Ishiguro, T.~Kai, T.~Kobayashi, Y.~Koga, and H.~Otsuka, {\em {Classification of Modular Symmetries in Type IIB Flux Landscape}}, \href{https://dx.doi.org/10.1093/ptep/ptaf112}{PTEP {\bfseries 2025} (2025) 093B07} {\ttfamily [\href{https://arxiv.org/abs/2502.20743}{arXiv:2502.20743}]}.

\bibitem{AtacamaCosmologyTelescope:2025nti}
{\bfseries Atacama Cosmology Telescope} Collaboration, {\em {The Atacama Cosmology Telescope: DR6 constraints on extended cosmological models}}, \href{https://dx.doi.org/10.1088/1475-7516/2025/11/063}{JCAP {\bfseries 11} (2025) 063} {\ttfamily [\href{https://arxiv.org/abs/2503.14454}{arXiv:2503.14454}]}.

\bibitem{Kawasaki:2023mjm}
M.~Kawasaki and T.~T.~Yanagida, {\em {Dai-Freed anomaly in the standard model and topological inflation}}, \href{https://dx.doi.org/10.1007/JHEP11(2023)106}{JHEP {\bfseries 11} (2023) 106} {\ttfamily [\href{https://arxiv.org/abs/2304.10100}{arXiv:2304.10100}]}.

\bibitem{Fukugita:1986hr}
M.~Fukugita and T.~Yanagida, {\em {Baryogenesis Without Grand Unification}}, \href{https://dx.doi.org/10.1016/0370-2693(86)91126-3}{Phys.\  Lett.\  B {\bfseries 174} (1986) 45--47}.

\bibitem{Asaka:1999yd}
T.~Asaka, K.~Hamaguchi, M.~Kawasaki, and T.~Yanagida, {\em {Leptogenesis in inflaton decay}}, \href{https://dx.doi.org/10.1016/S0370-2693(99)01020-5}{Phys.\  Lett.\  B {\bfseries 464} (1999) 12--18} {\ttfamily [\href{https://arxiv.org/abs/hep-ph/9906366}{hep-ph/9906366}]}.

\bibitem{Asaka:1999jb}
T.~Asaka, K.~Hamaguchi, M.~Kawasaki, and T.~Yanagida, {\em {Leptogenesis in inflationary universe}}, \href{https://dx.doi.org/10.1103/PhysRevD.61.083512}{Phys.\  Rev.\  D {\bfseries 61} (2000) 083512} {\ttfamily [\href{https://arxiv.org/abs/hep-ph/9907559}{hep-ph/9907559}]}.

\bibitem{Endo:2006qk}
M.~Endo, M.~Kawasaki, F.~Takahashi, and T.~T.~Yanagida, {\em {Inflaton decay through supergravity effects}}, \href{https://dx.doi.org/10.1016/j.physletb.2006.09.044}{Phys.\  Lett.\  B {\bfseries 642} (2006) 518--524} {\ttfamily [\href{https://arxiv.org/abs/hep-ph/0607170}{hep-ph/0607170}]}.

\bibitem{Endo:2007sz}
M.~Endo, F.~Takahashi, and T.~T.~Yanagida, {\em {Inflaton Decay in Supergravity}}, \href{https://dx.doi.org/10.1103/PhysRevD.76.083509}{Phys.\  Rev.\  D {\bfseries 76} (2007) 083509} {\ttfamily [\href{https://arxiv.org/abs/0706.0986}{arXiv:0706.0986}]}.

\bibitem{Endo:2006zj}
M.~Endo, K.~Hamaguchi, and F.~Takahashi, {\em {Moduli-induced gravitino problem}}, \href{https://dx.doi.org/10.1103/PhysRevLett.96.211301}{Phys.\  Rev.\  Lett.\  {\bfseries 96} (2006) 211301} {\ttfamily [\href{https://arxiv.org/abs/hep-ph/0602061}{hep-ph/0602061}]}.

\bibitem{Nakamura:2006uc}
S.~Nakamura and M.~Yamaguchi, {\em {Gravitino production from heavy moduli decay and cosmological moduli problem revived}}, \href{https://dx.doi.org/10.1016/j.physletb.2006.05.078}{Phys.\  Lett.\  B {\bfseries 638} (2006) 389--395} {\ttfamily [\href{https://arxiv.org/abs/hep-ph/0602081}{hep-ph/0602081}]}.

\bibitem{Kawasaki:2006gs}
M.~Kawasaki, F.~Takahashi, and T.~T.~Yanagida, {\em {Gravitino overproduction in inflaton decay}}, \href{https://dx.doi.org/10.1016/j.physletb.2006.05.037}{Phys.\  Lett.\  B {\bfseries 638} (2006) 8--12} {\ttfamily [\href{https://arxiv.org/abs/hep-ph/0603265}{hep-ph/0603265}]}.

\end{thebibliography}\endgroup

\end{document}